\documentstyle[12pt]{article}
\begin{document}
\begin{flushright}
Preprint IHEP 95-9\\
hep-ph/9602???\\
To appear in Phys.Lett.B
\end{flushright}

\begin{center}
{\large \bf
$f_B^{stat}$ and $\mu_\pi^2$ in quasiclassical approximation of
sum rules}\\
\vspace*{0.5cm}
V.V.Kiselev\\
{\it
Institute for High Energy Physics\\
Protvino, Moscow Region, 142284, Russia\\
E-mail: vkisselev@vxcern.cern.ch
}
\end{center}

\begin{abstract}
In the framework of sum rules with a use of quarkonium mass 
spectrum, evaluated
in the quasiclassical approximation, estimates of leptonic 
constant  $f_B^{stat}\simeq 320\pm 60$ MeV in a static limit 
and for the average heavy 
quark momentum squared $\mu_\pi^2 \simeq 0.5\pm 0.1$ GeV$^2$ 
are obtained.
\end{abstract}

{\bf 1.} In the Heavy Quark Effective Theory \cite{1}, used
for the description of strong interaction dynamics of heavy quarks, 
there are
some dimensionful parameters, which determine an accuracy of the
leading approximation in infinitely heavy quark limit 
as well as values of
power corrections over $\Lambda/m_Q \ll 1$, where $\Lambda$ is a scale,
determining the heavy quark virtuality inside hadrons. Among such 
parameters
in physics of heavy mesons $(Q\bar q)$ with a single heavy quark,
the most important quantities are the difference between masses of meson
and heavy quark $\bar \Lambda = M(Q\bar q) - m_Q$, the leptonic constant
of heavy meson $f_Q^{stat}$ in the static limit $m_Q \to \infty$, and
the square of heavy quark momentum $\mu_\pi^2$ inside the meson. Since
those values are determined by QCD at large distances, for estimates
one uses nonperturbative approaches, among which the most powerful tool is
sum rules \cite{2}.

As for the $\bar \Lambda$ value, its estimates in the framework of
QCD sum rules have been obtained in refs.\cite{3,4,4-}, where
$\bar \Lambda = 0.57\pm 0.07$ GeV. Moreover, the "optical"
sum rule by Voloshin \cite{5} allows one to get the inequality \cite{6}
\begin{equation}
\bar \Lambda > 2 \delta_1 \big(\rho^2-\frac{1}{4}\big) 
\simeq 0.59 \; \mbox{GeV}\;,
\end{equation}
where $\rho^2$ is the slope of universal Isgur--Wise function \cite{7},
and $\delta_1$ is the difference between the masses of the lightest
vector $S$-wave state and $P$-wave state for $(Q\bar q)$ system at
$m_Q \to \infty$.

Further, estimates of $f_B^{stat}$ in the framework of 
QCD sum rules and in
lattice computations are in agreement with each other and result in
\cite{1}
\begin{equation}
f_B^{stat} = 240 \pm 40 \; \mbox{MeV}\;.
\end{equation}
The sum rule estimation of average square of the heavy quark momentum
inside the meson gives the value \cite{1,4-,8}
\begin{equation}
\mu_\pi^2 = 0.5\pm 0.1 \; \mbox{GeV}^2\;,
\end{equation}
and the inequality \cite{6}
\begin{equation}
\mu_\pi^2 > 3\delta_1^2 \big(\rho^2-\frac{1}{4}\big)
\simeq 0.45 \; \mbox{GeV}^2\;.
\end{equation}
Note, however, that the values of the parameters $\delta_1$ and $\rho^2$
are presently rather uncertain, so that bounds (1) and (4) are not the most
conservative ones. A special discussion of the $\mu_\pi^2$ value can be
found, for instance, in ref.\cite{1}, where the role of a field theory
analog for the virial theorem is considered.

In the present letter we consider the QCD sum rules with a use of $S$-wave
level mass spectrum, calculated in the quasiclassical approximation, 
\cite{9,10,10-}
and obtain estimates of the $f_B^{stat}$ and $\mu_\pi^2$ values, which agree
with the results, given above.\\
\vspace*{3mm}

{\bf 2.} In recent papers \cite{9,10,10-} the QCD sum rules for leptonic
constants of $S$-wave levels in the $(Q_1\bar Q_2)$ quarkonium have been
considered with the use of the state mass spectrum, calculated in the 
quasiclassical approximation. For the $1S$-level one has got the expression
\begin{equation}
f^2_{V,P}\cdot M = \frac{16 \alpha_s}{\pi}\; \mu_\pi^2\; \mu\; H_{V,P}\;,
\label{7}
\end{equation}
where $\mu = m_1m_2/(m_1+m_2)$ is the reduced mass of quarkonium,
$\mu_\pi^2 = 2\mu \langle T\rangle$ is the average square of quark momentum
inside the quarkonium with the mass $M\simeq m_1+m_2$. 
$\alpha_s$ in eq.(\ref{7}) is evaluated at the scale of average virtuality of 
the one-gluon exchange between quarks, so 
$\alpha_s =\alpha_s^V(\sqrt{2}\mu_\pi)$ 
in the so-called $V$ scheme \cite{blm},
where $\Lambda^V_{QCD} = e^{5/6} \Lambda_{QCD}^{\overline{\rm MS}}$.
The $H_{V,P}$ factor corresponds to the hard gluon correction to
the vector and pseudoscalar currents, respectively, \cite{10-,sv,bra}
$$
H_{V,P} = 1 + \frac{2\alpha_s^H}{\pi}\biggl(\frac{m_2-m_1}{m_2+m_1}
\ln\frac{m_2}{m_1} - \delta_{V,P}\biggr)\;,
$$
where
$$
\delta_V=\frac{8}{3}\;,\;\;\;\delta_P = 2\;,
$$
and $\alpha_s^H$ is estimated at the scale $\mu_H= e^{3/8} m_Q$ 
in the $V$-scheme, if $m_Q=m_1=m_2$ \cite{19}.
The $H_{V,P}$ factors for the quark-to-antiquark annihilation currents
differ from the hard gluon correction to the quark-to-quark transition
currents \cite{sv}. Nevertheless, one can obtain the exact results for
$H_{V,P}$ from the factors, calculated in ref.\cite{sv}, by the symbolic
substitutions $V\to P$ and $m_1\to -m_1$ with the absolute value for the
logarithm argument \cite{10-}. However, this simple rule 
is not valid for the
scales, determining the coupling constant. For the vector and
axial-vector quark-to-quark transition currents, Neubert found \cite{16-}
$$
\mu_V = \sqrt{m_1 m_2} \exp\bigg\{\frac{3}{4}\bigg\}\;,\;\;\;\;
\mu_A = \sqrt{m_1 m_2} 
\exp\bigg\{\frac{2-5f(m_2/m_1)}{8-12f(m_2/m_1)}\bigg\}\;,
$$
with
$$
f(z)= \frac{1+z}{1-z} \ln{\frac{1}{z}} - 2\;.
$$
One can see, that at $m_1=m_2$ one has $\mu_H \neq \mu_{V,A}$.

Note, in the broad region of average distances between quarks: 
0.1 fm $< $ r $<$ 1 fm,
where the coulomb-like potential of heavy quark is transformed  into the 
linearly rising confining potential, the average kinetic energy 
$\langle T\rangle$ is a constant value, independent of $\mu$ 
(i.e. flavours), \cite{11,12} 
\begin{equation}
\langle T\rangle = \; \mbox{const.}
\end{equation}
This leads to that in the mentioned region of distances, the heavy 
quark potential is close to the logarithmic one \cite{12}, and
the quantization by the Bohr--Sommerfeld procedure results in
\begin{equation}
\frac{dM_n}{dn} = \frac{2\langle T\rangle}{n}\;.\label{9}
\end{equation}
In accordance with eq.(\ref{9}) and from spectroscopic data on the 
charmonium and bottomonium \cite{13}, one can get the estimate
\begin{equation}
\langle T\rangle =0.43\pm 0.01 \; \mbox{GeV.}
\end{equation}
However, the polynomial interpolation of masses for the excited states
in heavy quarkonia and heavy mesons\footnote{$(Q\bar q)$ masses are in 
agreement with estimates in potential models.} leads to the value
\begin{equation}
\langle T\rangle =0.38\pm 0.01 \; \mbox{GeV,}
\end{equation}
that is closer to the corresponding parameter of the logarithmic potential
\cite{12}. Therefore, in the following estimates we use the value
\begin{equation}
\langle T\rangle =0.40\pm 0.03 \; \mbox{GeV.}
\end{equation}
Note, that the approximate 
flavour-independence of the level spacing in heavy
quarkonia is the experimental observation, that can be reformulated in the
framework of phenomenological potential models, giving compact formulae for
the excitation energies, used as input parameters, fitted in the models.
A special simplification of the level spacing expressions appears in the
quasiclassical approximation, described above.

In the case of a heavy quarkonium $(Q\bar Q)$ with a hidden flavour
one has $4\mu = M$ and
\begin{equation}
\frac{f^2_{V,P}}{M} = \frac{2 \alpha_s}{\pi}\;\langle T\rangle \; H_{V,P}
\simeq \; \mbox{const.}, \label{11}
\end{equation}
where one can neglect the variation of $\alpha_s H_{V,P}$ value
under the heavy quark mass change \cite{10-}. Moreover, one has
$f_P\simeq f_V$ within the 5\% accuracy. Relation (\ref{11}) is
in a good agreement with experimental values of leptonic constants
for $\psi$- and $\Upsilon$-particles \cite{9,10}.

Since the threshold of the hadronic continuum in the system with
two heavy quarks is determined by masses of heavy mesons $(Q_1\bar q)$
and $(\bar Q_2 q)$, one finds \cite{14}
\begin{equation}
\bar \Lambda = \langle T\rangle \ln n_{th} \simeq
0.6\pm 0.1 \; \mbox{GeV,}
\end{equation}
where $n_{th}$ is the number of $S$-levels of heavy quarkonium below the
threshold of hadronic continuum ($n_{th}(b\bar b) =4$),
so that the estimation error is, in general, due to the variation of
$n_{th}$, $\delta \bar \Lambda = \langle T\rangle \delta n_{th}/ n_{th}
\simeq 0.1$ GeV.

For a heavy meson $(Q\bar q)$ a motion of the light current quark
in a medium of quark-gluonic condensate plays an essential role.
Therefore the most consistent consideration of sum rules requires the
use of the operator product expansion for quark currents with the account
of vacuum expectation values for operators of higher dimensions. However,
one can make the reasonable approximation and consider the case, when the
condensate influence generally results 
in the appearance of an effective mass
for the light quark. Such constituent quark can be considered as the 
nonrelativistic object, moving in the potential of static heavy 
quark\footnote{This approximation means that the "brown muck" is considered 
as a whole, i.e. with no internal structure.}. 
So, the potential quark models are quite successful in the heavy meson 
spectroscopy (see, for example, ref.\cite{GI}). Further, 
one can consider the phenomenological expressions, where
one does not include condensates, since the latters are implicitly
taken into account by means of  the introduction of some phenomenological
parameters such as the constituent mass.

Within the offered approach,
the approximation means that
$$
\mu \simeq M(Q\bar q) - m_Q = \bar \Lambda\;.
$$
 The introduction of the constituent light quark is the additional, 
but reasonable assumption to QCD or HQET, of course. It is an analog to a
nonperturbative quantity $E_c$, defining the thershold energy of hadronic
continuum in the HQET Laplace sum rules \cite{1,4-}. 
The $E_c$ value is determined
by the stability principle for the calculated parameters such as the
leptonic constant, say. The connection of $E_c$ to the quark-meson
mass gap is discussed in \cite{1}. The uncertainty of the $E_c$ estimation
in HQET is of the same order as that of in the constituent light quark mass.
In the finite energy sum rules, consistent with the HQET sum rules, the
$E_c$ value is the basic quantity, determining different dimensionful 
parameters (see ref.\cite{4-}, where explicit formulae are given). Thus, 
the $\mu$ value, determined by $\bar \Lambda$, has a quite enough accuracy,
comparable with the uncertainty in the other approaches.

Then one has
\begin{equation}
\mu_\pi^2 \simeq 2 \mu\; \langle T\rangle \simeq 0.5\pm 0.1
\; \mbox{GeV}^2\;.\label{13}
\end{equation}
In ref.\cite{15} one has shown that spectroscopic
data on the $\psi$- and $\Upsilon$-families give
\begin{equation}
\alpha_s(\psi, \Upsilon) H_{V,P} \biggl(\frac{2m_Q}{M}\biggr)^2 = 
0.21\pm 0.01\;, \label{13a}
\end{equation}
that is in agreement with the theoretical estimates \cite{10-}.
Using $\alpha_s^{\overline{\rm MS}}(m_Z) = 0.117\pm 0005$ \cite{13} as the
one-loop value, expressed in the form
$$
\alpha_s(m) = \frac{2\pi}{\beta_0(n_f)\ln{m/\Lambda^{(n_f)}}}\;,
\;\;\; \beta_0(n_f) = 11-\frac{2}{3}n_f\;,
$$
one finds $\Lambda^{(5)} = 85\pm 25$ MeV, so that $n_f$ is the number of
quark flavours with $m_{n_f} < m$. We use the one-loop rule for the
$\Lambda^{(n_f)}$ determination
$$
\Lambda^{(n_f)} = \Lambda^{(n_f+1)} 
\biggl(\frac{m_{n_f+1}}{\Lambda^{(n_f+1)}}
\biggr)^{2/(3\beta_0(n_f))}\;,
$$
leading to $\Lambda^{(3)} = 140\pm 40$ MeV 
and $\Lambda^{(4)} = 117\pm 30$ MeV.
Further, one takes $m_Q= M(Q\bar q)-\bar \Lambda$, 
so that $m_b=m_5=4.7\pm 0.1$
GeV, $m_c=m_4=1.4\pm 0.1$ GeV. One finds
\begin{equation}
\alpha_s^{\overline{\rm MS}}(m_b) = 0.20\pm 0.02\;,\label{13b}
\end{equation}
that agrees with $\alpha_s$ estimates\footnote{
The recent result by M.Voloshin gives 
$\alpha_s^{\overline{\rm MS}}(m_b)=0.185\pm 0.003$ \cite{19}.}
~ from experimental values of
the leptonic and radiative decay branching fractions for
$\Upsilon$ \cite{16} as well as with lattice computations for the
$(b\bar b)$ system spectroscopy \cite{17}, where the estimate, close to 
(\ref{13b}), takes place, too. 

Next, the factor of hard gluon correction is equal to
$$
H_P= 1.02\pm 0.01\;.
$$
It can be represented as the leading order 
approximation of the renormalization
group improved expression\footnote{
 Note, that $\mu$ is not the renormalization point, as one could think,
looking at eq.(\ref{rgi}). Hence, the $H_{V,P}$ factors do not contain an
explicit renormalization point dependence, which has to cancel against
the renormalization point dependence of some other parameters.}
\begin{equation}
H_{V,P}^{RG} = \biggl(\frac{\alpha_s(e^{\delta_{V,P}}\mu)}
{\alpha_s(m_Q)}\biggr)^{4/\beta_0(n_f)}\;,\label{rgi}
\end{equation}
that is known in HQET \cite{1} at $\delta_{V,P}=0$. Using eq.(\ref{rgi}),
one finds
$$
H_P^{RG}= 1.008\pm 0.004\;.
$$
So, we use
$$
H_P= 1.010\pm 0.005\;,
$$
that gives
\begin{equation}
\alpha_s^V(\sqrt{2}\mu_\pi)\; H_P = 0.36\pm 0.10\;. 
\end{equation}
Then in accordance with eq.(\ref{7}) one has
\begin{equation}
f_B^{stat} = 320 \pm 60 \; \mbox{MeV}\;.\label{15}
\end{equation}
\vspace*{3mm}

{\bf 3.} Thus, in the framework of sum rules with the use of
quarkonium spectroscopy, considered in the quasiclassical approximation,
one finds the estimates of $f_B^{stat}$ and $\mu_\pi^2$, which agree
with the values, obtained in QCD sum rules for the heavy meson currents.

As one can see, the obtained estimates of $f_B^{stat}$ and $\mu_\pi^2$
practically are near the bounds, derived in the sum rules \cite{5,6}.\\
\vspace*{3mm}

This work is partially supported by the ISF grants NJQ000, NJQ300 and 
the Program "Russian State Stipends for Young Scientists". Author expresses
special thanks to prof. V.Obraztsov for his hospitality at DELPHI, CERN,
where this work has been done.


\begin{thebibliography}{**}
\bibitem{1}
M.Neubert, Phys.Rep. 245 (1994) 259.
\bibitem{2}
M.A.Shifman, A.I.Vainshtein, V.I.Zakharov, 
Nucl.Phys. B147 (1979) 385, 448;\\
L.J.Reinders, H.Rubinstein, T.Yazaki, Phys.Rep. 127 (1985) 1.
\bibitem{3}
E.Bagan, P.Ball, V.Braun and H.Dosch, Phys.Lett. B278 (1992) 457.
\bibitem{4}
M.Neubert, Phys.Rev. D46 (1992) 1076.
\bibitem{4-}
S.Narison, Preprint CERN-TH.7549/94, 1994.
\bibitem{5}
M.Voloshin, Phys.Rev. D46 (1992) 3062.
\bibitem{6}
I.Bigi, A.G.Grozin, M.Shifman, N.G.Uraltsev, 
A.Vainshtein, Phys.Lett. B339
(1994) 160.
\bibitem{7}
N.Isgur and M.B.Wise, Phys.Lett. B232 (1989) 113, B237 (1990) 527.
\bibitem{8}
P.Ball and V.Braun, Phys.Rev. D49 (1994) 2472.
\bibitem{9}
V.V.Kiselev, Nucl.Phys. B406 (1993) 340.
\bibitem{10}
V.V.Kiselev, Preprint IHEP 94-63, Protvino, 1994, submitted to
Zh.Exp.Teor.Fiz. [JETP], hep-ph/9406243.
\bibitem{10-}
V.V.Kiselev, Preprint IHEP 95-63, Protvino, 1995, hep-ph/9504313.
\bibitem{blm}
S.J.Brodsky, G.P.Lepage and P.B.Mackenzie, Phys.Rev. D28 (1983) 228.
\bibitem{sv}
M.B.Voloshin, M.A.Shifman, Sov.J.Nucl.Phys. 47 (1988) 511.
\bibitem{bra}
E.Braaten, S.Fleming, Preprint NUHEP-TH-95-1, 1995, hep-ph/9501296.
\bibitem{19}
M.Voloshin, Univ. of Minnesota Preprint UMN-TH-1326-95, 1995.
\bibitem{16-}
M.Neubert, Phys.Lett. B341 (1995) 367.
\bibitem{11}
E.Eichten et al., Phys.Rev. D21 (1980) 203;\\
E.Eichten, Preprint FERMILAB-Conf-85/29-T, 1985;\\
A.Martin, Phys.Lett. 93B (1980) 338.
\bibitem{12}
C.Quigg and J.L.Rosner, Phys.Lett. B71 (1977) 153.
\bibitem{13}
L.Montanet et al., PDG, Phys.Rev. D50 (1994) 1173.
\bibitem{14}
V.V.Kiselev, Pisma v Zh.Exp.Teor.Fiz. 60 (1994) 498, 
[JETP Lett. 60 (1994) 509], hep-ph/9409256.
\bibitem{GI}
S.Godfrey, N.Isgur, Phys.Rev. D32 (1985) 189.
\bibitem{15}
V.V.Kiselev, Preprint IHEP 94-92, Protvino, 1994, hep-ph/9409288.
\bibitem{16}
T.Appelquist, H.D.Politzer, Phys.Rev.Lett. 34 (1975) 43;\\
A.De Rujula, S.L.Glashow, Phys.Rev.Lett. 34 (1975) 46;\\
V.A.Novikov et al., Phys.Rep. 41C (1978) 1.
\bibitem{17}
C.T.H.Davies et al., Florida State Univ. Preprint FSU-SCRI-94-79, 1994.
\end{thebibliography}
\end{document}